\documentclass[%
 reprint,
 groupedaddress,
 showpacs,showkeys,preprintnumbers,
 nofootinbib,
 nobibnotes,
 amsmath,amssymb,
 aps,
 prd,
 longbibliography,
 lengthcheck,%
]{revtex4-1}

\usepackage{graphicx}
\usepackage{dcolumn}
\usepackage{bm}
\usepackage{hyperref}

\begin{document}

\title{New type of parametrizations for parton distributions}

\author{A. Yu.~Illarionov}
\affiliation{Dipartimento di Fisica, Universita' di Trento, Italy}
\email{illario@science.unitn.it}

\author{A. V.~Kotikov}
\affiliation{Joint Institute for Nuclear Research, Russia}

\author{S. S.~Parzycki}
\affiliation{Joint Institute for Nuclear Research, Russia}

\author{D. V.~Peshekhonov}
\affiliation{Joint Institute for Nuclear Research, Russia}

\date{\today}

\begin{abstract}
New type of parametrization for parton distribution functions,
based on an exact their $Q^2$-evolution at large and small $x$ values, is
constructed for valence quarks. It preserves exactly 
the low $x$ and large $x$ asymptotics of the solution of the DGLAP equation
and obeys the Gross-Llewellyn-Smith 
sum rule.
\end{abstract}



\maketitle

\section{ Introduction }

\begin{figure}[htb]
\begin{center}
\includegraphics[width=0.5\textwidth,clip]{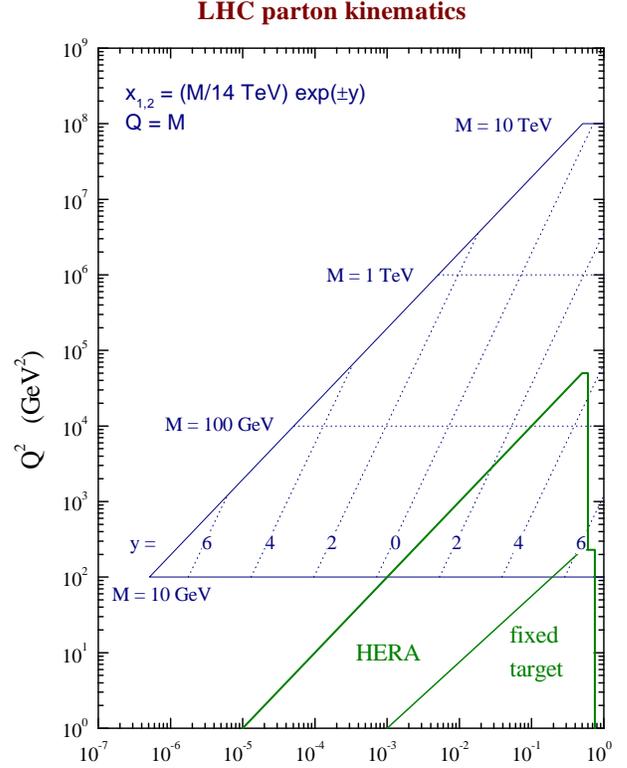}
\vspace{0.6cm}
\caption{%
The $(x,Q^2)$ ranges of the PDFs
contributed to the LHC processes and
ones fitted at HERA and fixed target experiments
(from \cite{Thorne2005}).
}%
\end{center}
\label{LHCkin}
\end{figure}

The parton distribution functions (PDFs) which
contribute to the LHC processes, and the PDFs
fitted at HERA and fixed target experiments
are defined at somewhat
different $(x,Q^2)$ ranges
\footnote{Several recent PDF fits, 
as well as the references to the previous
studies, can be found in \cite{NNLOfits}.}
(see, for example, Fig.~\ref{LHCkin} taken from \cite{Thorne2005}).
Therefore, a direct application
of the modern PDF sets \cite{NNLOfits} to the LHC processes
is not well justified.

This problem is really important, because the
larger uncertainties for many processes at the LHC
originate mainly from our restricted knowledge 
of the parton distributions 
(see, for example, the recent paper \cite{Forte} and references therein).
 
In the present paper we propose
an idea to overcome this problem. Our solution consists of
the two basic steps.
Firstly, we find asymptotics of solutions of
the Dokshitzer-Gribov-Lipatov-Altarelli-Parisi (DGLAP)
equations \cite{DGLAP} for the parton densities at low and large $x$ values of the
Bjorken variable and, at the next step,
we approach a combination of the two solutions for the full range of $x$.

In a sense, this is not a new idea. A similar approach
had been given by the Spanish group (see book \cite{Yndu} and references therein) 
about 40 years ago. However, in the present paper the 
parametrization will be constructed in a rather
different way. In particular,
it includes
an important subasymptotic term which
is fixed exactly by the sum rules.

The results of the present paper are restricted by the leading order (LO) of the 
perturbation expansion,
what is reasonable \cite{Sherstnev}, 
since for many processes
at the LHC the next-to-leading-order (NLO) corrections are not known so far.
Moreover, in the present study
we limit ourselves
to the valence
quarks, whose evolution does not contain
contributions from the gluons.

\section{ A short theoretical input }


In this section we briefly present the theoretical part of our analysis.
The reader is referred to \cite{Kotikov2007} for more details.

The deep-inelastic scattering (DIS)
$l + N \to l^\prime + X$, where
$l$ and $N$ are 
the incoming lepton and nucleon, and $l^\prime$ is the outgoing lepton,
in the one of basic processes to study of the nucleon structure. The DIS cross-section
can be split to the lepton
$L^{\mu \nu}$ and hadron $F^{\mu \nu}$ parts
\begin{equation}
d\sigma \sim L^{\mu \nu} F^{\mu \nu} \, .
\label{S1.1}
\end{equation}

The lepton part $L^{\mu \nu}$
is evaluated exactly, while the hadron one,
$F^{\mu \nu}$, can be presented in the following form
\begin{eqnarray}
F^{\mu \nu} &=& \left(-g^{\mu \nu}+ \frac{q^{\mu}q^{\nu}}{q^2}\right) \, F_1(x,Q^2)
\nonumber \\
&+& \left(p^{\mu}+ \frac{(pq)}{q^2}q^{\mu}\right) \left(p^{\nu}+
\frac{(pq)}{q^2}q^{\nu}\right) \, F_2(x,Q^2)
\nonumber \\
&+& i\varepsilon_{\mu \nu \alpha \beta}p^{\alpha}p^{\beta} \frac{x}{q^2}
\, F_3(x,Q^2) + \dots \,,
\label{S1.2}
\end{eqnarray}
where the symbol $\dots$ stands for the parts, which depend on 
the nucleon spin.
The functions
$F_k(x,Q^2)$ (hereafter $k=1,2,3$) are the DIS structure functions (SFs) and
$q$ and $p$ are the photon and parton
momenta. Moreover, the two variables
\begin{equation}
Q^2=-q^2 > 0 ~,~~ x=\frac{Q^2}{2(pq)}
\label{S1.var}
\end{equation}
determine the basic properties of the DIS process. Here, $Q^2$ is the ``mass''
of the virtual photon and/or $Z/W$ boson, and the Bjorken variable $x$
($0<x<1$) is the part of the hadron momentum carried by the scattering parton (quark or gluon).

\subsection{Mellin transform}

The Mellin transform
diagonalizes the $Q^2$ evolution of the parton densities.
In other words, 
the $Q^2$ evolution of the Mellin moment with certain value
$n$ does not depend on the moment with another value $n^\prime$.

The Mellin moments $M_k(n,Q^2)$ of the SF $F_k(x,Q^2)$
\begin{equation}
M_k(n,Q^2) = \int_{0}^{1}dx\,\,x^{n-1} F_k(x,Q^2)
\label{S1.3}
\end{equation}
can be represented
as the sum
\begin{equation}
M_k(n,Q^2) = \sum_{a=q,\bar{q},G} \, C^a_k(n,Q^2/\mu^2) \,
A_a(n,\mu^2),
\label{S1.4}
\end{equation}
where $C^a_k(n,Q^2/\mu^2)$ are the coefficient functions and
$A_a(n,\mu^2)=<N| \mathcal{O}^a_{\mu_1, ..., \mu_n}|N>$ are the matrix
elements of the Wilson operators $\mathcal{O}^a_{\mu_1, ..., \mu_n}$,
which in turn are process independent.

Phenomenologically, the matrix elements $A_a(n,\mu^2)$ are equal to
the Mellin moments of the PDFs $f_a(x,\mu^2)$,
\footnote{
All parton densities are multiplied by $x$, i.e. 
in the LO the
structure functions are some combinations of the parton densities.}
where $f_a(x,\mu^2)$ 
are the parton distributions of quarks ($a=q_i$), antiquarks ($a=\bar{q}_i$) with
$(i=1, \cdots ,6$),
 and gluons ($a=G$),
i.e.
\begin{equation}
A_a(n,\mu^2) \equiv f_a(n,\mu^2) = \int_{0}^{1}dx\,\,x^{n-1} f_a(x,\mu^2) \, .
\label{S1.5}
\end{equation}
The coefficient functions $C^a_k(n,Q^2/\mu^2)$ are represented by 
\begin{equation}
C^a_k(n,Q^2/\mu^2) = \int_{0}^{1}dx\,\,x^{n-1} \tilde{C}^a_k(x,Q^2/\mu^2)
\label{S1.6}
\end{equation}
and responsible for the relationship between SFs and PDFs. Indeed, in
the $x$-space the relation (\ref{S1.4}) is replaced by
\begin{equation}
F_k(x,Q^2) = \sum_{a=q,\bar{q},G} \,
\tilde{C}^a_k(x,Q^2/\mu^2) \otimes f_a(x,\mu^2) \, ,
\label{S1.7}
\end{equation}
where $\otimes$ denotes the Mellin convolution
\begin{equation}
f_1(x) \otimes f_2(x) \equiv \int^1_x \, \frac{dy}{y} \, f_1(y)
f_2\left(\frac{x}{y}\right) \, .
\label{S1.8}
\end{equation}

Using Eqs.~(\ref{S1.4}) and (\ref{S1.7}),
one can fit
the shapes of PDFs $f_a(x,\mu^2)$, which are process-independent and
use them later on for another processes. Indeed, the
cross-sections of the hadron-hadron
processes are proportional
to
the parton luminosities $f_{a,b}(x,\mu^2)$
(see, for example, the recent paper \cite{Neubert} and references therein),
which
are the Mellin convolutions of the
two PDFs $f_{a}(x,\mu^2)$ and $f_{b}(x,\mu^2)$:
\begin{equation}
f_{a,b}(x,\mu^2) = f_a(x,\mu^2) \otimes f_b(x,\mu^2) \, .
\label{S1.9}
\end{equation}


\subsection{Quark distribution functions}

The distributions of the $u$ and $d$ quarks contain the valence 
and the sea parts
\begin{equation}
f_{q_1} \equiv f_u = f_u^V + f_u^S ~,~~ f_{q_2} \equiv f_d = f_d^V + f_d^S \, .
\label{S1.10}
\end{equation}
The distributions of the other quark flavors and of
all the antiquarks contain the sea parts only:
\begin{equation}
f_{q_j} = f_{q_j}^S,~~(j=3,4,5,6),~~~ f_{\bar{q}_i} = f_{\bar{q}_i}^S ~~(i=1, \cdots ,6)\, .
\label{S1.11}
\end{equation}

It is useful to define the combinations of quark densities,
the valence part $f_{V}$, the sea one $f_{S}$ and the singlet one $f_{SI}$
\cite{Buras}
\footnote{
Here
we consider
all quark flavors.
Really, heavy quarks factorize out when $\sqrt{Q^2}$ becomes
less then their masses, and we should exclude them from the $Q^2$-region.
}
\begin{eqnarray}
f_{V} &=& f_u^V + f_d^V ~,~~
f_{S} = \sum_{i=1}^{6} \left(f_{q_i}^S + f_{\bar{q}_i}^S \right) ~,
\nonumber \\
f_{SI} &=& \sum_{i=1}^{6} \left(f_{q_i} + f_{\bar{q}_i} \right) = f_{V} + f_{S} \, .
\label{S1.12}
\end{eqnarray}

Because the PDFs, which contribute to the structure functions, 
are accompanied by some numerical factors, there are also nonsinglet parts
\begin{equation}
f_{\Delta_{ij}} = \left(f_{q_i} +f_{\bar{q}_i}\right)-
 \left(f_{q_j} +f_{\bar{q}_j}\right)\, ,
\label{S1.13}
\end{equation}
which contain difference of densities of quarks and antiquarks with
different values of charges.

As an example, we consider the electron-proton scattering, where the
corresponding SF has the form
\begin{equation}
F_2^{ep}(x,Q^2) = \sum_{i=1}^{6} e_i^2  \left(f_{q_i}(x,Q^2) +
f_{\bar{q}_i}(x,Q^2) \right) \, .
\label{S1.14}
\end{equation}
In the four-quark case (when $b$ and $t$ quarks are separated out),
as in \cite{Buras}, we will have
\begin{equation}
F_2^{ep}(x,Q^2) = \frac{5}{18} \, f_{SI}(x,Q^2) + \frac{1}{6} f_{\Delta}(x,Q^2) \, ,
\label{S1.15}
\end{equation}
where
\begin{align}
f_{\Delta} &= \sum_{q_i=u,c} \left(f_{q_i}(x,Q^2) +
f_{\bar{q}_i}(x,Q^2) \right)
\nonumber \\
&- \sum_{q_i=d,s} \left(f_{q_i}(x,Q^2) +
f_{\bar{q}_i}(x,Q^2) \right)
\, .
\label{S1.16}
\end{align}

\subsection{DGLAP equation}

The PDFs obey the DGLAP equation \cite{DGLAP}
\begin{equation}
\frac{d}{d\ln {\mu^{2}}} \, f_{a}(x,\mu^{2}) ~=
\sum_{b} \tilde{\gamma}_{ab}(x) \otimes f_{b}(x,\mu^{2}) \,,
\label{S2.1}
\end{equation}
where $a,b=NS,SI,G$ and $\tilde{\gamma}_{ab}(x)$ are the so-called
splitting functions.

Anomalous dimensions $\gamma_{ab}(n)$ of the twist-two Wilson operators
$\mathcal{O}^a_{\mu_1, ..., \mu_n}$ in the brackets $b$
are the Mellin transforms of the corresponding
splitting functions
\begin{equation}
\gamma _{ab}(n) = \int_{0}^{1}dx\,\,x^{n-1} \tilde{\gamma}_{ab}(x) \,.
\label{S2.2}
\end{equation}

In the Mellin moment space, the DGLAP equation becomes to be
the standard renormalization-group equation
\begin{equation}
\frac{d}{d\ln {\mu^{2}}} \, f_{a}(n,\mu^{2}) ~=
\sum_{b=NS,SI,G}\tilde{\gamma}_{ab}(n)\, f_{b}(n,\mu^{2}) \,.
\label{S2.2a}
\end{equation}

Below we will study the properties of the valence part only.
Consideration of the other quark densities will be discussed separately.

\section{Low and large $x$ asymptotics}


The large $x$ asymptotics of the valence quark density has
the following form \cite{Gross,LoYn}
\begin{equation}
f_V(x,Q^2) ~\to~ B_V(s) (1-x)^{\beta_V(s)}  \,,
\label{S2.3}
\end{equation}
where
\begin{gather}
s=\ln \left(\frac{\ln (Q_0^2/\Lambda^2)}{\ln (Q^2/\Lambda^2)}\right) \,,~
\beta_V(s) = \beta_V(0) + \hat{d}_{V} s \,,~
\hat{d}_{V} = \frac{16}{3\beta_0} \,,
\nonumber \\
B_V(s) =
B_V(0) \frac{e^{-p_V s}}{\Gamma(1+\beta_V(s))} \,,~
p_V =
\hat{d}_{V} \left(\gamma_E-\frac{3}{4}\right) s \,.
\label{S2.4}
\end{gather}
$B_V(0)$ and $\beta_V(0)$ are free parameters.
Here $\beta_0=11-2f/3$ is the first term of the QCD $\beta$-function,
$f$ is the number of active quarks
and $\gamma_E$ is the Euler constant. The constant $\beta_V(0)$ can be estimated
from the quark counting rules \cite{schot} as
\begin{equation}
\beta_V(0)
\sim 3 \,.
\label{S2.5}
\end{equation}
Eqs.~(\ref{S2.3}) and (\ref{S2.4}) demonstrate the fall
of the parton densities at large $x$ values when $Q^2$ increases.


At small-$x$ values the valence part
has the following asymptotics  \cite{FMartin,LoYn,Kotikov1993}
\begin{equation}
f_V(x) ~\to~  A_V(s) \, x^{\lambda_V} \,,
\label{S3.1}
\end{equation}
where
\begin{gather}
A_V(s) = A_V(0) e^{-d_{NS}(1-\lambda_V) s} \,,
\label{S3.2} \\
d_{NS}(n) =
\frac{16}{3\beta_0} \left[
\Psi(n+1)+ \gamma_E -\frac{3}{4}- \frac{1}{2n(n+1)} \right] \,.
\nonumber
\end{gather}
$\lambda_V$ and $A_V(0)$ are free parameters
and $\Psi(n+1)$ is Euler $\Psi$-function.

From the Regge calculus
the constant $\lambda_V \sim 0.3 \div 0.5$.
Moreover,
the $Q^2$ evolution of this parton density shows that
$\lambda_V$ should be $Q^2$ independent \cite{Kotikov1996}.

In Eq.~(\ref{S3.2}) the ``anomalous dimension'' $d_{NS}(n)$ is represented in
the form which is useful for the non-integer $n$ values. 
Usually, the elements of
the coefficient functions and anomalous dimensions are expressed as the combinations
of the nested sums \cite{Vermaseren}, which can be expanded, however, to the
non-integer $n$ values according to \cite{KaKo}.

\section{Parametrization
}

The valence quark part $f_V(x,Q^2)$ can be represented in the following form
\footnote{
Following to \cite{Maximov}, it is possible to add to the parametrization
(\ref{S4.1}) an additional polynomial term
$(1+ \sum^N_{k=1}\alpha_{V,k} x^k)$ with certain constants
$\alpha_{V,k}$. This term fixes properly the PDF
shape and may improve essentially an
agreement with the corresponding experimental data.
In the valence sector, considered here,
we found a good agreement between (\ref{S4.1}) and the data without such term.
However, in the singlet part, where there is a strong
correlation between the see quark and gluon densities, similar terms may be
useful.}
\begin{equation}
f_V(x,Q^2) =  x^{\lambda_V} (1-x)^{\beta_V(s)}
\biggl[A_V(s)(1-x) + B_V(s)x  \biggr] \,,
\label{S4.1}
\end{equation}
which is constructed as a combination of the small $x$ and large $x$
asymptotics,
the last term is equal to $A_V(s)$ at small $x$ and to
$B_V(s)$ at large $x$ values.
The $Q^2$-dependence of the parameters in (\ref{S4.1}) is given by Eqs.~(\ref{S2.4})
and (\ref{S3.2}).

\subsection{Gross-Llewellyn-Smith sum rule}

The additional relation between the parameters in (\ref{S4.1}) 
stems from the LO Gross-Llewellyn-Smith 
sum rule \cite{GrLl}
\footnote{Above LO, the Gross-Llewellyn-Smith sum rule \cite{GrLl} is
defined as the integral of the structure function $F_3$ and contains
the perturbative ($\sim \alpha_s$) and the power corrections in its r.h.s.
(see, for example, \cite{KaSi} and references therein).}

\begin{equation}
\int_0^1 \frac{dx}{x} f_V(x,Q^2) = Q_V ~,~~ Q_V=3 \,,
\label{S4.2}
\end{equation}
which simply informs about the number of the valence quarks in nucleon.

As long as the Eq.~(\ref{S4.1}) is just a parametrization, an attempt to apply
the sum rule for, e.g., $k$ different values of $Q^2$ produces $k$
additional relations that is, of course, unacceptable.
The sum rule can be applied only at one point $s=s_C$ for 
some \emph{critical} value of $Q^2_C$.
For the other values of $Q^2$, the sum rule holds only approximately
and one can estimate the deviation from the exact sum rule.

So, we have the following relation
\begin{align}
Q_V &= \frac{\Gamma(\lambda_V)\Gamma(1+\beta_V(s_C))}{\Gamma(\lambda_V+2+\beta_V(s_C))}
\Bigl[
\lambda_V B_V(0) e^{-p_V s_C}
\nonumber\\
&+ (1+\beta_V(s_C)) A_V(0) e^{-d_{NS}(1-\lambda_V) s_C}
\Bigr] \,.
\label{S4.3}
\end{align}

It is possible to choose any ``middle'' value of $s$. However, if
$s_C=0$, i.e. $Q^2_C=Q^2_0$,
then Eq.~(\ref{S4.3}) is considerably simplified
\begin{equation}
Q_V = \frac{\Gamma(\lambda_V)\Gamma(1+\beta_V(0))}{\Gamma(\lambda_V+2+\beta_V(0))}
\Bigl[
\lambda_V B_V(0)
+ (1+\beta_V(0))A_V(0)
\Bigr] \,.
\label{S4.4}
\end{equation}

\subsection{Subasymptotic term }

In the present paper we choose another possibility
to take exactly the sum rule (\ref{S4.2}) into account. We introduce
an additional term $\sim D_V(s)$ to our parameterization (\ref{S4.1}), which
can be written as 
\begin{align}
f_V(x,&Q^2) = x^{\lambda_V} (1-x)^{\beta_V(s)}
\label{S4.5}\\
&\times \Bigl[A_V(s)(1-x) + B_V(s)x + D_V(s)\sqrt{x}(1-x) \Bigr] \,,
\nonumber
\end{align}
where the last term in the brackets is subasymptotic in both the small
and the large $x$ values.
These types of the subasymptotic reduction,
$\sim \sqrt{x}$ at small $x$
and $\sim (1-x)$ at large $x$ values, have been
discussed in \cite{Yndu}.

Now, the sum rule (\ref{S4.2}) can be satisfied at any $Q^2$ values and
determines completely the new term,
\begin{eqnarray}
Q_V &=& \frac{\Gamma(\lambda_V)\Gamma(1+\beta_V(s))}{\Gamma(\lambda_V+2+\beta_V(s))}
\, \Bigl[ \lambda_V B_V(s)  + (1+\beta_V(s)) A_V(s) \Bigr]
\nonumber \\
&+&
\frac{\Gamma(\lambda_V+1/2)\Gamma(2+\beta_V(s))}{\Gamma(\lambda_V+5/2+\beta_V(s))}
\, D_V(s) \,.
\label{S4.6}
\end{eqnarray}

\section{Results}


To obtain the parameters of our parameterization (\ref{S4.5}) for the
valence part, we compare it numerically with the results of several
Gluck-Reya-Vogt (GRV) sets \cite{GRV92,GRV98} and
Gluck-Jimenez-Delgado-Reya (GJR) one \cite{GRV08}.
The choice of the GRV/GJR evolution is related to future
investigations of the gluon and sea quark densities; the small $x$
asymptotics obtained in \cite{Q2evo}
are conceptually close to the GRV/GJR approach.

\begin{table*}[tH]
\begin{ruledtabular}
\begin{center}
\begin{tabular}{|c|ccccccc|}
    & $f$ & $\alpha_s(M_Z)$ & $Q_0^2~(\mbox{GeV}^2)$ & $A_V(0)$ & $B_V(0)$ & $\lambda_V$ & $\beta_V(0)$ \\ \hline
GRV(1991) \cite{GRV92} & 4 & 0.1282 & 0.25 & 0.61 & 80.62 & 0.317 & 2.870 \\
GRV(1998) \cite{GRV98} & 3 & 0.1250 & 0.26 & 0.93 & 64.80 & 0.333 & 2.812 \\
GJR(2008) \cite{GRV08} & 3 & 0.1263 & 0.30 & 1.07 & 85.60 & 0.390 & 2.945 \\ 
\end{tabular}
\end{center}
\caption{%
The parameters of the valent quark distribution of Eq.~(\ref{S4.5}) fitted on
several GRV/GJR sets.
}%
\label{Table:param}
\end{ruledtabular}
\end{table*}



%

%

With the parameters, reported in the Table~\ref{Table:param}
our parameterization (\ref{S4.5}) and the GRV/GJR sets
are in good agreement for all $x$ values in the very broad $Q^2$ range,
$0.25$~GeV$^2$ $<Q^2<10^4$~GeV$^2$.
\footnote{We used $Q^2$ evolution at fixed $f$ value. The contributions
of the heavy-quark thresholds are negligible \cite{KK09}.}

Then, the parameterization (\ref{S4.5}) describes the experimental data as
well as the GRV/GJR sets and has the analytic form containing the exact
asymptotics (\ref{S2.3}) and (\ref{S3.1}) of the
DGLAP
evolutions at small and large $x$ values. Thus, it can be applied with 
good warranty in any other ($x,Q^2$) region, where the 
experimental data are still not available. 
So, it should be applicable for the LHC range of $x$ and $Q^2$ values.

Note, that all GRV/GJR sets themselves give quite close results for the valence quarks,
i.e. these results should not change significantly when the new
data will appear, for example, from the LHC experiments.


\begin{figure}[htb]
\begin{center}
\includegraphics[width=0.47\textwidth,clip]{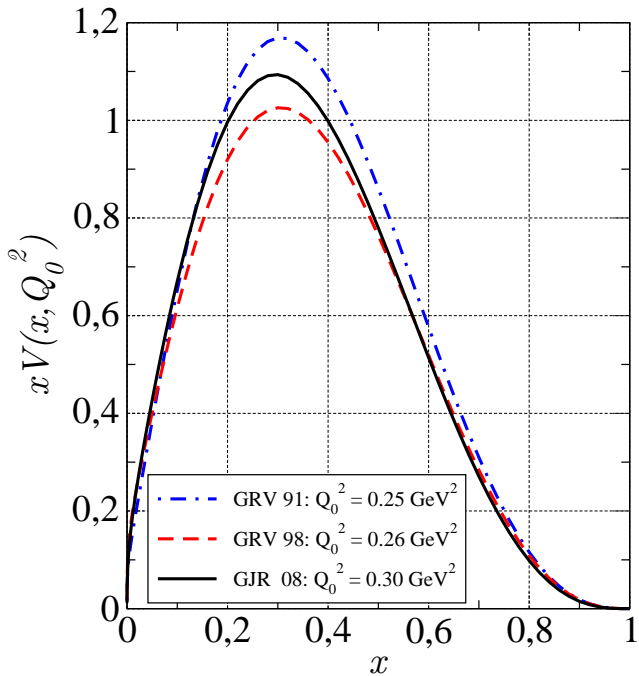}
\caption{%
The valence quark density at $Q^2_0$.
}%
\end{center}
\label{v_Q02}
\end{figure}

\begin{figure}[htb]
\begin{center}
\includegraphics[width=0.47\textwidth,clip]{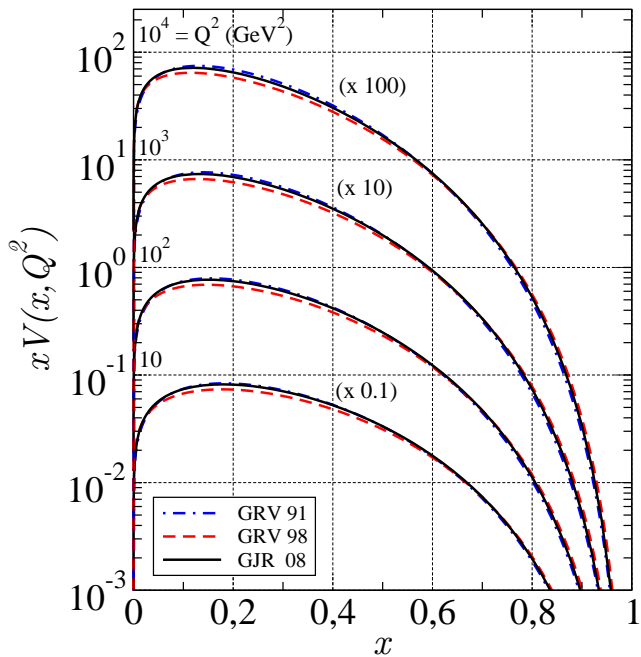}
\caption{%
The valence quark density at several
values of $Q^2$: $Q^2=10^m$ GeV$^2$ with $m=1,2,3$ and $4$.
}%
\end{center}
\label{valent}
\end{figure}

\section{Conclusions}

In this work, we investigated
the low $x$ and large $x$ asymptotics of the 
valence quark density, and
obtained the corresponding parametrization (\ref{S4.5}).
%
This parametrization obeys explicitly the Gross-Llewellyn-Smith sum rule. 
It has been compared with various GRV/GJR sets
\cite{GRV92,GRV98,GRV08}
in order to fit the initial values of
all the $Q^2$-dependent parameters.
It has been performed accurately, because
numerically the form (\ref{S4.5}) and the GRV/GJR sets have very similar
shapes for all $x$ and $Q^2$ values.

At the next step, we plan to add to our analysis the nonsinglet and sea
quark densities as well as
the gluon distribution,
and
apply the obtained results to the
analysis of several LHC processes. Moreover, a comparison of these results
with
the predictions of other sets \cite{NNLOfits} of the parton 
densities will be done.
We also plan to consider the PDFs in nuclei. Therefore,
the EMC effect \cite{EMC}, which is very important in the high-energy regime,
will be added to the consideration.

\section{Acknowledgements}
We are grateful to Zaza Merebashvili and Igor Cherednikov for careful reading the text and to
the authors of \cite{GRV08} for the correspondence.
The work was supported by RFBR grant No.10-02-01259-a.

Calculations were partially performed on the HPC facility of SISSA/Democritos in Trieste
and partially on the HPC facility ``WIGLAF'' of the Department of Physics,
University of Trento.

\end{document}